\documentclass[sigconf]{acmart}

\usepackage{amsmath,amsfonts}
\usepackage[utf8]{inputenc} 
\usepackage{algorithmic}
\usepackage{graphicx }
\usepackage{textcomp}
\usepackage{xcolor}
\usepackage{balance}
\usepackage{mathrsfs}

\usepackage{siunitx}

\usepackage{subcaption}
\usepackage{caption}

\newcommand{\xth}[1]{{\textit{#1}}\mbox{-th}}

\newtheorem{remark}{Remark}

\usepackage{multirow}
\usepackage{booktabs}
\usepackage{array}

\AtBeginDocument{%
  }

\setcopyright{acmlicensed}
\copyrightyear{2018}
\acmYear{2018}
\acmDOI{XXXXXXX.XXXXXXX}

\acmISBN{978-1-4503-XXXX-X/18/06}

\begin{document}

\title{Enhanced Drug Delivery via Localization-Enabled Relaying in Molecular Communication Nanonetworks}

\author{Ethungshan Shitiri}
\affiliation{%
 \institution{Universitat Politècnica de Catalunya,}
  \city{Barcelona}
  \country{Spain}
}

\author{Akarsh Yadav}
\affiliation{%
  \institution{Purdue University,}
  \city{Indiana}
  \country{USA}
}

\author{Sergi Abadal}
\affiliation{%
 \institution{Universitat Politècnica de Catalunya,}
  \city{Barcelona}
  \country{Spain}
}

\author{Eduard Alarcón}
\affiliation{%
 \institution{Universitat Politècnica de Catalunya,}
  \city{Barcelona}
  \country{Spain}
}

\author{Ho-Shin Cho}
\affiliation{%
  \institution{Kyungpook National University}
  \city{Daegu}
  \country{South Korea}
}

\renewcommand{\shortauthors}{Shitri et al.}

\begin{abstract}
Intra-body nanonetworks hold promise for advancing targeted drug delivery (TDD) systems through molecular communications (MC). In the baseline MC-TDD system, drug-loaded nanomachines (DgNs) are positioned near the infected tissues to deliver drug molecules directly. To mitigate the decline in drug delivery efficiency caused by diffusion, we propose an enhanced MC-TDD system with a relay network. This network employs a novel localization-enabled relaying mechanism, where a nano-controller broadcasts a localization signal. DgNs then measure the received signal strength against thresholds to determine their clusters relative to the infected tissue. Additionally, our study considers the effect of multiple absorbing DgNs on the channel impulse response (CIR), a factor overlooked in previous works. Our approach improves drug delivery efficiency by $17\%$ compared to the baseline system. Importantly, we find that optimizing CIR is crucial for enhancing drug delivery efficiency. These findings pave the way for further research into optimizing CIR-based relay selection, as well as investigating the impact of factors such as drug molecule lifespan, obstruction probabilities, and flow dynamics.
\end{abstract}

\keywords{intra-body, molecular communication, nanonetworks, relay, targeted drug delivery, localization}

\acmYear{2024}\copyrightyear{2024}
\setcopyright{acmlicensed}
\acmConference[NanoCom '24]{International Conference on Nanoscale Computing and Communication}{October 28--30, 2024}{Milan, Italy}
\acmBooktitle{International Conference on Nanoscale Computing and Communication (NanoCom '24), October 28--30, 2024, Milan, Italy}
\acmDOI{10.1145/3686015.3689347}
\acmISBN{979-8-4007-1171-8/24/10}
\maketitle

\section{Introduction}\label{sec_Introduction}
Nano-scale communications are gaining widespread interest due to their potential to interconnect nanomachines, forming nanonetworks \cite{akyildiz2008nanonetworks}. These nanonetworks are particularly promising for enabling intra-body nanonetworks. Molecular communication (MC), which transmits and receives information via molecules, is promising for intra-body nanonetworks due to its nanoscale operation and inherent biocompatibility. In contrast, Terahertz (THz) waves, another promising candidate for nanonetworks, face significant challenges, including spreading, molecular absorption, and scattering losses in bodily fluids and tissues. Additionally, the biocompatibility of THz waves remains under investigation, necessitating further research to ensure their long-term safety \cite{Llatser2012GrapheneAnt}. Enabling implantable devices with communication capabilities without compromising biocompatibility is a key feature for advanced healthcare systems \cite{Felicetti2016Applications}. This drives the exploration of MC in various innovative intra-body healthcare applications \cite{nakano2005molecular, Felicetti2016Applications, Chude2017SurveyMCTDDD, Shitiri2021Timing, Guo2021}.

\begin{figure}
\centering \includegraphics[width=\columnwidth,keepaspectratio]{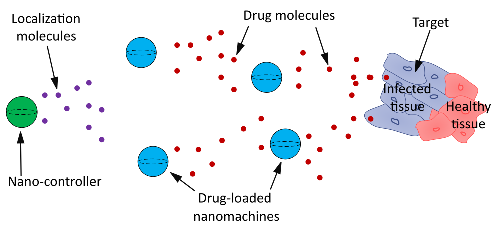}
\caption{Schematic of the proposed MC-TDD system with a nano-controller and random DgN placement.}
\label{fig_system_model_basic_MCTDD} \vspace{-.3cm}
\end{figure}

Among the various applications, MC-based targeted drug delivery (MC-TDD) systems can benefit significantly from MC as it allows for coordinated drug release among drug-loaded nanomachines (DgNs), which are tiny devices capable of storing and releasing drug molecules. This enhances autonomy and reduces dependency on external control. MC-TDD systems, however, present unique challenges. In the baseline MC-TDD system, DgNs release drug molecules directly to the infected tissue. Due to the random nature of diffusion, some molecules stray from their target, decreasing treatment efficacy and potentially causing side effects in healthy tissues \cite{Guengerich2011Toxicity,xia2020DrugEfficacy}. Therefore, it is crucial to improve the delivery efficiency of drug molecules while minimizing stray molecules to enhance the effectiveness of these systems \cite{tewabe2021TDD}.

Recent studies have shown that introducing relays can significantly enhance MC system performance by facilitating the transmission of drug molecules over shorter distances \cite{Wang2015Relay, Cheng2022JointOptimizations}. However, these studies do not adequately address scenarios where relays are randomly distributed. While this is a significant challenge in practical applications, establishing an effective relay network under these conditions requires accurate localization of DgNs. Traditional localization techniques like GPS are impractical for DgNs due to their high energy consumption and inefficiency. Hence, there is a need for novel localization solutions specific to MC-TDD systems.

To address these challenges, we propose an enhanced MC-TDD system that incorporates a relay network of DgNs. As illustrated in Fig. \ref{fig_system_model_basic_MCTDD}, this system features a nano-controller alongside the DgNs. The nano-controller is the system's central coordinator, managing both the localization and other communication processes among DgNs and with external devices. DgNs are capable of absorbing and forwarding drug molecules. In addition, we develop a low-complexity localization mechanism to form a relay network among the DgNs. Briefly, the nano controller sends out this localization signal; DgNs measure the strength of the signal and classify themselves into clusters. Intermediate clusters capture stray drug molecules and relay them to the infected tissue, creating efficient delivery links. The relay DgNs forward drug molecules without amplification, thereby conserving energy. This method is both power- and computationally-efficient, eliminating the need for complex position estimation procedures. To our knowledge, this is the first study to investigate MC-TDD systems with multiple relays and their impact.

The remainder of the paper is organized as follows. Section \ref{sec_system} discusses the envisioned MC-TDD system setup, including the preliminaries: multiple receiver model, channel impulse response, and received drug molecules. In Section \ref{sec_proposed}, we introduce the proposed method and describe its operation in detail. In Section \ref{sec_numerical_results}, we present the numerical results, and in Section \ref{sec_conclusion}, we provide discussions and conclusions. Table \ref{table_notations} lists the commonly used notations and their definitions.

\begin{table}[]
\caption{Notations}
\begin{tabular}{|p{1cm}|p{6cm}|}
\hline
Notation        & Definition                                                \\ \hline
$G$             & Drug-carrying nanomachine                                 \\ \hline
$C$             & Nano-controller                                           \\ \hline
$S$             & Infected tissue                                           \\ \hline
$N$             & Number of relay hops                                      \\ \hline
$\mathcal{K}$   & Set of DgNs. $|\mathcal{K}| = K$                      \\ \hline
$\mathcal{K}_n$ & Set of DgNs in cluster $n, n=0,1,\ldots$. $|\mathcal{K}_n| = K_n$\\ \hline
$\mathcal{K}_m$ & Set of DgNs in relay cluster, $m > n$. $|\mathcal{K}_m| = K_m$   \\ \hline
$G_{k}$         & \xth{k} DgN                                               \\ \hline
$G_{k,n}$       & \xth{k} DgN in the \xth{n} cluster                        \\ \hline
$G_{j,m}$       & \xth{j} DgN in the \xth{j} cluster                        \\ \hline
$d_{Tx,Rx}$     & Distance between $Tx \in \{C,G_{k,n}\}$ and $Rx \in \{G_{k},G_{j,m}, S\}$ \\ \hline
$\mathbf{d}_{{\mathcal{K}_n},Rx}$     & Set of distances between DgNs in cluster $\mathcal{K}_n$ and $Rx \in \{G_{j,m}, S\}$ \\ \hline
$L$             & Number of timeslots                                      \\ \hline
$\Delta t$      & Sampling interval                                        \\ \hline
$L_\text{sp}$   & Number of sampling intervals                             \\ 
\hline

$N_z^\text{TX}$& Number of molecules released. $z\in \{\text{loc, drug}\}$ \\ \hline
$N_{z,i}^\text{RX}$& Number of molecules received during the \xth{i} timeslot. $z\in \{\text{SX, MX}\}$ \\ \hline

\end{tabular} \label{table_notations}
\end{table}

\begin{figure}
    \centering    \includegraphics[width=\columnwidth,keepaspectratio]{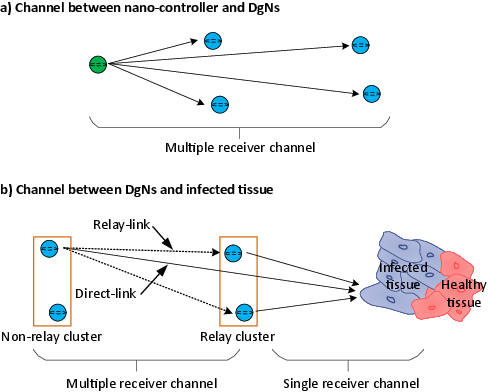}
    \caption{Channel models between a) the nano-controller and DgNs and b) the DgNs and infected tissue.}
    \label{fig_system_model} \vspace{-.3cm}
\end{figure} 

\section{System Model} \label{sec_system}
In the proposed MC-TDD system, \textit{cf.} Fig. \ref{fig_system_model_basic_MCTDD}, the position of the nano-controller before the DgNs is a design choice to maintain unidirectional signaling to ensure that communication flows efficiently from the nano-controller to the DgNs and ultimately to the infected tissue. This setup facilitates a streamlined communication process, enabling potential extensions of the proposed system to flow-assisted channels. In this study, the following are the considerations for transmitter (Tx) and receiver (Rx).

\textbf{Nano-controller ($C$): } It acts as a Tx as it releases the localization molecules. 

\textbf{DgN ($G$):} A DgN can function as both Rx and Tx, depending on whether it is absorbing localization/drug molecules or releasing drug molecules. 

\textbf{Infected tissue ($S$):}
The infected tissue acts as the final Rx, absorbing drug molecules transmitted by the DgNs. 

\subsection{Channel}\label{sec_channel}
We consider a three-dimensional unbounded channel without medium flow, such as the extracellular matrix of the connective tissue \cite{Femminella2015MCDrugDel}. The Tx is modeled as a point source, which is valid when the distance to the Rx exceeds the Rx’s radius \cite{Noel2016ChannelIR}. The Rx is modeled as a spherical, fully absorbing body that counts the absorbed molecules over defined observation periods \cite{Jamali2019ChannelTut}. The localization and drug molecules are assumed to have the same radius and, consequently, the same diffusion coefficient, but they are distinct and distinguishable from each other \cite{Kim2013NovelMod}.

The channel can be classified as a single or multiple-receiver channel, as shown in Fig. \ref{fig_system_model}.

\textbf{Single-receiver (SX) Channel:} When only one Rx is present, typically the infected tissue.

\textbf{Multiple-receiver (MX) Channel:} When multiple Rxs, such as multiple DgNs or combinations of DgNs and infected tissue, are present.

\subsection{Channel Impulse Response}\label{sec_CIR}
The channel impulse response (CIR) between a Tx and an Rx, separated by a distance $d_{Tx,Rx}$, is the probability that the Rx absorbs a molecule released by the Tx within a small time period $\Delta t$. Here, $\Delta t$ is the sampling time interval during which an observation is made to measure the number of absorbed molecules. The CIR can be expressed as \cite{Jamali2019ChannelTut}
\begin{equation} \label{eq_CIR}
h(d_{Tx,Rx}) = F(d_{Tx,Rx}) - F(d_{Tx,Rx}-\Delta t),
\end{equation}
where $F(\cdot)$ is the cumulative distribution function, which represents the probability the Rx will absorb a molecule until time $t$. For the SX case, $F(\cdot)$ is described by 
\begin{equation}
   F_\text{SX}(d_{Tx,Rx}) =  \int_0^t f_\text{SX}(d_{Tx,Rx},x)\,dx  = \frac{r_{Rx}}{d_{Tx,Rx}+r_{Rx}}\text{erfc}\left[\frac{d_{Tx,Rx}}{\sqrt{4Dt}}\right]  \label{eq_F_hit_SX}
\end{equation} and MX case by
\begin{equation}
F_\text{MX}(d_{Tx,Rx}) = \int_0^t f_\text{MX}(d_{Tx,Rx},x)\,dx, \label{eq_F_hit_MX}
\end{equation}
where $r_{Rx}$ denotes the radius of the receiver and $f(\cdot)$ denotes the probability density function of molecule absorption. For more details on the density functions, readers are directed to \cite{schulten2015lectures} for SX case and \cite{yaylali2021channelMx} for MX case. Note that $\int_0^t f_\text{MX}(d_{Tx,Rx},x)\,dx$ is intractable and is therefore solved numerically.

\subsection{Number of Received Molecules}
The number of received molecules received by a DgN or the infected tissue is random and can be modeled as a Binomial process \cite{Kuran2010Energy}. If the number of molecules released is large and $\Delta t\rightarrow 0$, the Binomial distribution can be approximated by the Normal distribution \cite{Shitiri2021Probability}. Let the time $t$ be divided into timeslots of length $T$, each of which are composed of the $L_{\text{sp}}$ sampling intervals. i.e., $\Delta t$, (\textit{cf.} Fig. \ref{fig_time_overview}). Accordingly, the number of received molecules by Rx during the \xth{i} timeslot, denoted by $N_i^\text{RX}(d_{Tx,Rx})$, can be expressed as 
        \begin{align} \label{eq_NormalApprox}
            N_i^\text{RX}(d_{Tx,Rx}) \sim \mathscr{N}\left(\mu_i(d_{Tx,Rx}),\,\sigma_i^2(d_{Tx,Rx}) \right),
        \end{align}
where $\mathscr{N}\left(x,y\right)$ represents the Normal distribution with mean $x$ and variance $y$. The mean $\mu_i(d_{Tx,Rx})$ is given by 
    \begin{align} \label{eq_Nrx}
        \mu_i(d_{Tx,Rx}) &=  N^\text{TX}\sum_{l=1}^{ L_\text{sp}}h(d_{Tx,Rx},(i-1)T+l \,\Delta t) \\ &\triangleq N^\text{TX}\, \mathcal{H}_i(d_{Tx,Rx}) ,
    \end{align}
and the variance $\sigma_i^2(d_{Tx,Rx})$ is expressed as 
        \begin{align} \label{eq_Var_NRx}
            \sigma_i^2(d_{Tx,Rx}) \triangleq N^\text{TX}\,  \mathcal{H}_i(d_{Tx,Rx})\big(1- \mathcal{H}_i(d_{Tx,Rx})\big),
        \end{align}
where $\mathcal{H}_i(d_{Tx,Rx})= \sum_{l=1}^{ L_\text{sp}}F(d_{Tx,Rx},(i-1)T+l\,\Delta t) -  F(d_{Tx,Rx},(i-1)T+(l-1)\,\Delta t)$.

\section{Proposed Localization-enabled Relaying Mechanism} \label{sec_proposed}
Building on the benefits of relay networks, our scheme enables DgNs to relay drug molecules efficiently without requiring explicit localization. This is achieved by leveraging the key property of multiple absorbing Rxs, where each Rx absorbs molecules, effectively turning molecule release by a Tx into a broadcast transmission. Accordingly, nearby Rxs predominantly absorb molecules, reducing the number available for distant Rxs. While this limits the communication range of a Tx, a common drawback in MC systems, we capitalize on it to develop our localization mechanism. We assume no path obstructions for simplicity, noting that such obstacles would degrade performance irrespective of the relay network. Nonetheless, their impact will be explored in future work.

\begin{figure}
    \centering    \includegraphics[width=0.9\columnwidth,keepaspectratio]{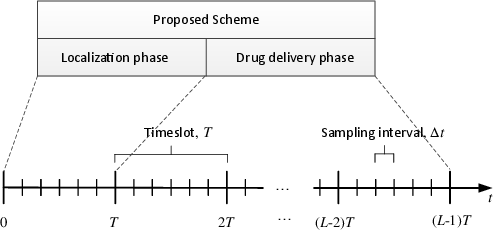}
    \caption{Overview of the proposed localization-enabled relaying scheme and timeslot model.} 
    \label{fig_time_overview}
\end{figure}

Following a \textit{localization phase}, the proposed scheme progresses to a \textit{drug delivery phase}, as shown in Fig. \ref{fig_time_overview}. This two-phase approach ensures proper organization of DgNs before drug delivery. For illustrative purposes, we assume $L$ timeslots, although the exact number will vary based on specific therapy requirements and designer discretion. The first timeslot is dedicated to localization, while the remaining $(L-1)$ timeslots are used for drug delivery. Notation-wise, $i=1$ denotes the localization phase, while $i\ge2, i \in [2,L]$ denotes the drug delivery phase.

\subsection{Localization phase operation} \label{sec_loc_phase_oper}
At the start of the localization phase ($t=0$), the nano-controller releases $N^\text{TX}_\text{loc}$ localization molecules. By the end of this phase ($t=T$), the DgNs evaluate the received localization molecules against predefined thresholds, $\eta_0,\eta_1,\ldots,\eta_{N-1}$, to determine their cluster. These thresholds are set in descending order ($\eta_0>\eta_1>\ldots>\eta_{N-1}$) because localization is performed relative to the infected tissue, not the nano-controller. For an $N$-hop relay network, this process results in the formation of $(N+1)$ clusters.

The number of localization molecules received by DgN $G_k$ at a distance $d_{C,G_k}$ from the nano-controller during the localization timeslot ($i=1$) is denoted as $N^\text{RX}_{\text{MX},1}(d_{C,G_k})$. Here, the MX channel model is applied since all DgNs can receive the localization molecules. Then, the localization operation can be represented as
    \begin{equation} \label{eq_local_timeslot}
        G_k \in
        \begin{cases}
               \mathcal{K}_0,& \text{if } N^\text{RX}_{\text{MX},1}(d_{C,G_k})>\eta_0, \quad \forall k\in \mathcal{K}\\
               \mathcal{K}_1,& \text{if }\eta_0>N^\text{RX}_{\text{MX},1}(d_{C,G_k})>\eta_1, \\
               \quad \vdots &\quad\quad \vdots\\
               \mathcal{K}_{N-1},& \text{if }\eta_{(N-2)}>N^\text{RX}_{\text{MX},1}(d_{C,G_k})>\eta_{N-1}, \\
               \mathcal{K}_N,& \text{if }N^\text{RX}_{\text{MX},1}(d_{C,G_k})<\eta_{N-1}.
        \end{cases}
    \end{equation}
Here $N^\text{RX}_{\text{MX},1}(d_{C,G_k})$ is derived using \eqref{eq_F_hit_MX}-\eqref{eq_Var_NRx}. 

\begin{remark}\label{remark1}
Cluster $\mathcal{K}_0$ includes DgNs likely furthest from the infected tissue and acts as a non-relay cluster. Conversely, cluster $\mathcal{K}_N$ comprises DgNs closest to the infected tissue, serving as the final hop where molecules from clusters  $\mathcal{K}_0$ to $\mathcal{K}_{N-1}$ are ultimately relayed to the infected tissue.
\end{remark}

\subsection{Drug delivery phase operation}\label{sec_drug_deliver_oper}
After the localization phase, the system transitions into the drug delivery phase. DgNs in each cluster release drug molecules simultaneously at the beginning of each drug delivery timeslot. The drug molecules can be delivered through two pathways: \textit{relay links}, where drug molecules are forwarded through subsequent clusters, and \textit{direct links}, where drug molecules are delivered directly to the infected tissue.

\begin{figure}
    \centering    \includegraphics[width=.95\columnwidth,keepaspectratio]{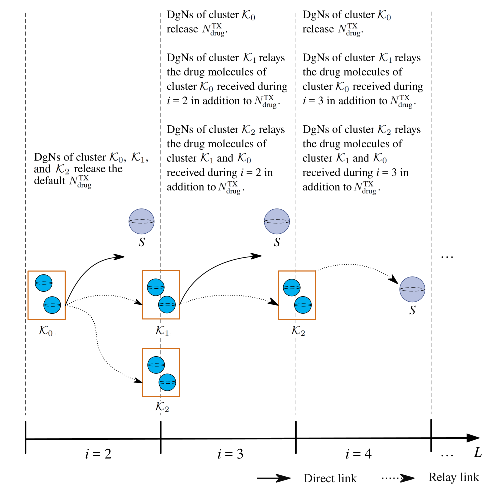}
    \caption{Timing diagram of the drug delivery operation from the DgNs in cluster $\mathcal{K}_{0}$.}
    \label{fig_relaytiming_operation}
\end{figure}

To illustrate this process, consider the timing diagram shown in Fig. \ref{fig_relaytiming_operation}, where we focus on the drug delivery process from cluster $\mathcal{K}_{0}$ in a three-cluster relay network.
At the start of the first drug delivery timeslot ($i=2$), each DgN in cluster $\mathcal{K}_{0}$ releases $N^\text{TX}_{\text{drug}}$, which is the \textit{default number of drug molecules} that each DgN is designed to release. Most of these molecules are absorbed by DgNs in cluster $\mathcal{K}_{1}$, while fewer molecules reach cluster $\mathcal{K}_{2}$ and the infected tissue $S$. In the subsequent timeslot $(i=3)$, the DgNs in clusters $\mathcal{K}_{1}$ release \textit{cumulative drug molecules}, which include those received from cluster $\mathcal{K}_{0}$ at $i=2$. (The cumulative drug release will be formally defined in \eqref{eq_NTx_relays}). Most of these molecules are then absorbed by cluster $\mathcal{K}_{2}$, with a small fraction reaching $S$. Continuing this process in timeslot $i=4$, DgNs in cluster $\mathcal{K}_{2}$ release the cumulative drug molecules, including those from cluster $\mathcal{K}_{0}$ and cluster $\mathcal{K}_{1}$, ultimately delivering them to $S$. 

In sum, in clusters \(\mathcal{K}_{n},\ n \neq N\), DgNs deliver the drug molecules to the infected site using both relay and direct links. While DgNs in cluster \(\mathcal{K}_{N}\) deliver drug molecules solely via direct links, as they are the closest to the infected tissue.

\begin{figure}
    \centering
    \includegraphics[width=.95\columnwidth,keepaspectratio]{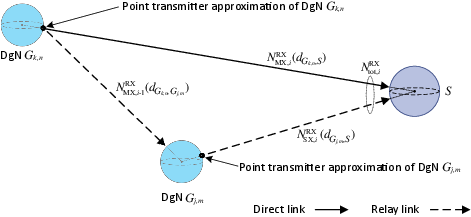}
    \caption{Illustration of the drug delivery operation in the presence of a relay.} 
    \label{fig_drug_delivery_operation}
\end{figure}

Next, we determine the number of drug molecules received by the DgNs and infected tissue. Given that the DgNs and the infected tissue can absorb drug molecules, the MX channel model applies here, as illustrated in Fig. \ref{fig_drug_delivery_operation}. For simplicity, we depict the channel model with one transmitting DgN, $G_{k,n}$, and a relay DgN, $G_{j,m}$.

Firstly, during the \xth{i} timeslot, DgN $G_{k,n}$ releases drug molecules, which are received by relay DgN $G_{j,m}$ and the infected tissue $S$. Then, the number of drug molecules received by DgN $G_{j,m}$ from DgNs in cluster $\mathcal{K}_{n}$ can be expressed as
    \begin{equation} \label{eq_relays_recv_mols}
        N_i^\text{RX}(\mathbf{d}_{\mathcal{K}_{n}, G_{j,m}})\triangleq \sum_{k=1}^{K_n}  N^\text{RX}_{\text{MX},i}(d_{G_{k,n},G_{j,m}}),
    \end{equation}
where $N^\text{RX}_{\text{MX},i}(d_{G_{k,n},G_{j,m}}) =N^\text{TX}_{\text{drug}} \mathcal{H}_i(d_{G_{k,n},G_{j,m}})$ is the number of drug molecules received by DgN $G_{j,m}$ from DgN $G_{k,n}$ that is at a distance of $d_{G_{k,n},G_{j,m}} \in \mathbf{d}_{\mathcal{K}_n,G_{j,m}}$ during the \xth{i} timeslot. From \eqref{eq_relays_recv_mols}, it can be inferred that the proposed scheme achieves full diversity because all the DgNs of cluster $\mathcal{K}_{m}$ can serve as relays for each DgNs of cluster $\mathcal{K}_{n}$.

\begin{table*}[t!]\centering
 \caption{Impact of varying $\eta$ on the CIR of direct links.}
\begin{tabular}{ccll}\toprule
$\eta_{0_1}$ & $\eta_{0_2}$ & $F_x(d_{k,S},t),x =$ \{SX,MX\}   & CIR at $\eta_{0_2}$        \\
\midrule
 $G_1 \in \mathcal{K}_{0}$   &  $G_1 \in \mathcal{K}_{0}$    & $F_\text{MX}(d_{G_1,S},t)$  & degrades                     \\
 $G_2 \in \mathcal{K}_0$   &  $G_2 \in \mathcal{K}_{1}$       & $ F_\text{MX}(d_{G_2,S},t) \rightarrow  F_\text{SX}(d_{G_2,S},t)$ & improves    \\
 $G_3 \in \mathcal{K}_1$      &  $G_3 \in \mathcal{K}_1$        & $F_\text{SX}(d_{G_3,S},t)$  & unchanged        \\
\bottomrule               
\end{tabular}\label{table_EffectsOfEtaonCIR} 
\end{table*} 

Similarly, the amount of drug molecules received by $S$ from DgNs in cluster $\mathcal{K}_{n}$ can be expressed as 
   \begin{equation} \label{eq_direct_path}
        N_{i}^\text{RX}(\mathbf{d}_{\mathcal{K}_n, S})\triangleq \sum_{k=1}^{K_n}  N^\text{RX}_{\text{MX},i}(d_{G_{k,n},S}),
    \end{equation}
where $N^\text{RX}_{\text{MX},i}(d_{G_{k,n}, S})=N^\text{TX}_{\text{drug}} \mathcal{H}_i(d_{G_{k,n},S})$ is the number of drug molecules received by the infected tissue from DgN $G_{k,n}$ that is at a distance of $d_{G_{k,n},S} \in \mathbf{d}_{\mathcal{K}_n,S}$ during the \xth{i} timeslot. In the absence of relays, $ N_{i}^\text{RX}(\mathbf{d}_{\mathcal{K}_n, S})$ will adopt the SX channel model, i.e., $N_{i}^\text{RX}(\mathbf{d}_{\mathcal{K}_n, S})\triangleq \sum_{l=1}^{\mathcal{K}_n} N^\text{RX}_{\text{SX},i}(d_{G_{k,n},S})$ and can be obtained via \eqref{eq_F_hit_SX}, \eqref{eq_CIR}-\eqref{eq_Var_NRx}. 

During the same timeslot, DgN $G_{j,m}$ also releases drug molecules to $S$. In this case, the infected tissue is the sole receiver. Hence, we can employ the SX channel model. Then, the number of drug molecules received by $S$ from the DgNs in cluster $\mathcal{K}_{N}$ during the \xth{i} timeslot can be expressed as 
   \begin{equation} \label{eq_direct_path_nearby}
        N_{i}^\text{RX}(\mathbf{d}_{\mathcal{K}_{N}, S})\triangleq \sum_{m=1}^{K_N}  N^\text{RX}_{\text{SX},i}(d_{G_{j,m},S}),
    \end{equation}
where $N^\text{RX}_{\text{SX},i}(d_{G_{j,m},S})=N^\text{TX}_{\text{drug},i,G_{j,m}} \mathcal{H}_i(d_{G_{j,m},S})$ is the number of drug molecules received by the infected tissue from DgN $G_{j,m}$ that is at a distance of $d_{G_{j,m},S} \in \mathbf{d}_{\mathcal{K}_{N},S}$ during the \xth{i} timeslot. $N^\text{TX}_{\text{drug},i,G_{j,m}}$ is the cumulative drug release that includes the drug molecules received during the \xth{(i-1)} timeslot along with $N^\text{TX}_{\text{drug}}$.

\subsection{Analysis of drug molecules delivered. } \label{sec_total_drugs}
$T$ is assumed to be sufficiently large to maximize the delivery of drug molecules within each timeslot. As demonstrated earlier, most drug molecules released are absorbed by nearby relay DgNs, which then forward the molecules toward the infected tissue over multiple timeslots. In an $N$-hop relay network, molecules released by DgNs in cluster $\mathcal{K}_{n}$ require $(N-n+1)$ timeslots to reach the infected tissue. Thus, the total time required for complete drug delivery, represented by the maximum number of timeslots, is $(N+1)$. This is the time required for molecules from cluster $\mathcal{K}_{0}$ to pass through all relay stages to the infected tissue. Therefore, to effectively evaluate the drug delivery performance, it is crucial to assess the number of drug molecules delivered at the $(N+1)$ timeslot and the subsequent timeslots, where each subsequent timeslot reflects continuous delivery from previous releases. Such timeslots can be expressed as
        \begin{equation} \label{eq_timslot_Index}
            i_\text{del} = i+N,\ i=2,3,\ldots.
        \end{equation}
        
Before deriving the total drug molecules delivered, we define the cumulative drug release. Recall that each DgN of the relay clusters $\mathcal{K}_{m},\forall m,$ performs a cumulative drug release during timeslot $i,\ i \ge 3$, that can vary during each of those timeslots. The cumulative drug release of DgN $G_{j,m}$ at timeslot $i,\ i \ge 3$, can be expressed as 
\begin{equation}\label{eq_NTx_relays}
    N^\text{TX}_{\text{drug},i,G_{j,m}} = N_{i-1}^\text{RX}(\mathbf{d}_{\mathcal{K}_n, G_{j,m}})+N^\text{TX}_{\text{drug}},\ i \ge 3, 
\end{equation}
where $N_{i-1}^\text{RX}(\mathbf{d}_{\mathcal{K}_n, G_{j,m}})$ is derived according to \eqref{eq_relays_recv_mols}. Then, for the cases when $3 \le i \le L$ and $\forall m$, $N^\text{TX}_{\text{drug},i,G_{j,m}}$ is applied instead of $N^\text{TX}_\text{drug}$. Note that in the case of no relays, $N^\text{TX}_{\text{drug},i,G_{j,m}}$ reduces to $N^\text{TX}_\text{drug}$.

Then, the total amount of drug delivered to the infected tissue during the $i_\text{del}$ timeslots can be expressed as
    \begin{align} \label{eq_tissue_recv_mols}
        N_{\text{tot},i_\text{del}}^\text{RX}&\triangleq \sum_{n=0}^N N_{i_\text{del}}^\text{RX}(\mathbf{d}_{\mathcal{K}_n, S}),\notag \\ 
        & = \sum_{k=1}^{K_0}N^\text{TX}_{\text{drug}}\mathcal{H}_{i_\text{del}}(d_{G_{k,0},S}) + \notag \\
        & \qquad \qquad \sum_{m=1}^N \sum_{j=1}^{K_m}N^\text{TX}_{\text{drug},i_\text{del},G_{j,m}}\mathcal{H}_{i_\text{del}}(d_{G_{j,m},S}).
    \end{align}
The term $\mathcal{H}_{i_\text{del}}(\cdot)$ is described by the MX channel model except for $\mathcal{H}_{i_\text{del}}(d_{G_{j,m},S}), m = N$, which is describe by the SX channel model.

\subsection{Analysis of drug delivery links. } \label{sec_drug_del_paths}
We define the total number of drug delivery links as the number of connections that facilitate the transportation of drug molecules from the DgNs to the infected tissue. In scenarios without relays, drug delivery completes within the same timeslot it was transmitted, simplifying the analysis to a single timeslot for drug delivery, i.e., $i_\text{del}=2,3,\ldots$. However, in the proposed system involving relays, the analysis must account for drug delivery through the relays, taking $(N+1)$ timeslots that span from drug delivery timeslots $i$ to $i_\text{del}$.

All DgNs maintain a direct link to the infected site regardless of the CIR, resulting in $K$ direct links. Without relays, the total number of drug delivery links equals $K$. When relays are used, each relay DgN serves as a relay for other DgNs. Thus, the number of relay links is given by 
\begin{align}
\sum_{n=0}^{N-1} \sum_{m=n+1}^{N} K_{n} \cdot K_{m}.
\end{align}
Therefore, the total number of drug delivery links with relays is 
\begin{align}\label{eq_tot_links}
K + \sum_{n=0}^{N-1} \sum_{m=n+1}^{N} K_{n} \cdot K_{m}.
\end{align}
For brevity, we shall refer to the total number of drug delivery links as total links.

\subsection{Analysis of threshold's impact on the CIR. } \label{sec_analysis_CIR_eta}
Let us consider a one-hop relay network, i.e., $N=1$, resulting in $2$ clusters, $\mathcal{K}_0$ and $\mathcal{K}_1,$ and one threshold, $\eta_0$. We consider three DgNs, namely, $G_1$, $G_2$, and $G_3$, ordered by their distance to the infected tissue, i.e., $d_{G_1,S} > d_{G_2,S} > d_{G_3,S}$. To illustrate the impact of $\eta$ on the CIR of direct links, we consider two values of $\eta_0$, namely $\eta_{0_1}$ and $\eta_{0_2}$, where $\eta_{0_1}< \eta_{0_2}$. 

As shown in Table \ref{table_EffectsOfEtaonCIR}, when the threshold is $\eta_{0_1}$, assume DgNs $G_1$ and $G_2$ are localized in cluster $\mathcal{K}_{0}$, while $G_3$ is localized in cluster $\mathcal{K}_{1}$. 
The channels of $G_1$ and $G_2$ can be described as a single Tx and MX case (infected tissue and DgN $G_3$) case. Meanwhile, $G_3$'s channel is a single Tx and SX case (infected tissue). At the higher threshold $\eta_{0_2}$, DgN $G_2$ is re-localized to cluster $\mathcal{K}_{1}$ because it cannot cross $\eta_{0_2}$. In this scenario, although the channel of $G_1$ is still an MX, its direct link's CIR degrades because the number of receivers increases by adding $G_2$ to cluster $\mathcal{K}_2$. Conversely, the channel of $G_2$ changes from MX to SX case, improving its direct link's CIR. Finally, the channel of $G_3$ remains the same; hence, its direct link's CIR remains unchanged.  

Therefore, it can be deduced that the CIR of:
\begin{itemize}
    \item direct links degrade with the number of relays and could become insignificantly low.
    \item relay links also degrade with the number of relays but could be relatively better than the CIR of direct links. Thus, having more relays will not necessarily yield a higher drug delivery performance, and an optimal number of relays is anticipated.
    \item DgNs of cluster $\mathcal{K}_N$ is independent of the number of relays and would generally have the best CIR.
\end{itemize}

\begin{figure} 
\centering
  \includegraphics[width=\columnwidth]{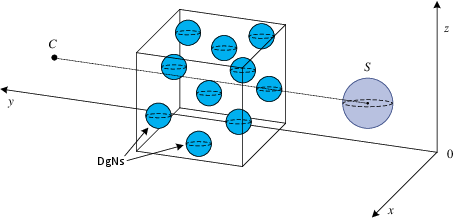} 
\caption{Deployment region of the DgNs and the locations of the nano-controller and infected tissue. (Not drawn to scale.)}  
\label{fig_deploy_region}
\end{figure}

\section{Numerical Results} \label{sec_numerical_results}
We evaluate the performance of the proposed protocol by testing its capability to establish a one-hop network as a representative scenario.
The DgNs are randomly located in a 3D region measuring $\SI{70}{\micro\meter}\times \SI{70}{\micro\meter}\times \SI{70}{\micro\meter}$ as shown in Fig. \ref{fig_deploy_region}. The center coordinate of the spherical infected tissue is $(\SI{30}{\micro\meter},\SI{8}{\micro\meter},\SI{30}{\micro\meter})$ and the point nano-controller coordinate is  $(\SI{30}{\micro\meter},\SI{106}{\micro\meter},\SI{30}{\micro\meter})$. The minimum distance between the nano-controller or infected tissue and the DgNs is $\SI{10}{\micro\meter}$ and the minimum distance between DgNs is $\SI{6}{\micro\meter}$. These minimum distance values align with established models in the literature, particularly those applying point transmitter approximations \cite{Oguzhan2022Multiple}. 
We set $L=3$; hence, the total drug molecules delivered are measured at the third timeslot, $N_{\text{tot},3}^\text{RX}$. For a one-hop network, two clusters will be formed, namely, $\mathcal{K}_\text{0}$ and $\mathcal{K}_\text{1}$. Lastly, for performance evaluations, we compare the results to the system without relays, considered the baseline. Table \ref{table_Sim_Param} lists the values of the remainder of the simulation parameters. The results shown here are the average values over $4000$ repetitions.

\begin{table}[t!]
\centering 
\caption{Simulation parameters}
\begin{tabular}{p{4.99cm}p{2.7cm}}
\toprule
Parameter                      & Values                   \\ \midrule
Diffusion coefficient, $D$                            & \SI{79.4}{\micro\metre^{2}\per\second} \cite{yilmaz2014Three}        \\
No. of DgNs, $K$                            & $10$             \\
Radius of DgNs                 & \SI{5}{\micro\metre}  \cite{yilmaz2014Three}              \\
Radius of infected tissue        & \SI{8}{\micro\metre}               \\
No. of Monte Carlo simulations & $4000$ \\ 

Timeslot duration, $T$ & \SI{5}{\second}\\ 
Sampling duration, $\Delta t$ & \SI{0.01}{\second}\\
No. of timeslots, $L$ & $3$\\
No. of drug molecules released, $N^\text{TX}_{\text{drug}}$ & $10000$ molecules\\
No. of localization molecules released, $N^\text{TX}_{\text{loc}}$ & $10000$ molecules\\
Localization threshold, $\eta$ & \SIrange{0}{1000}{} molecules\\
\bottomrule                 
\end{tabular}\label{table_Sim_Param}
\end{table}

\begin{figure}[t!] 
\centering
  \includegraphics[width=0.9\columnwidth]{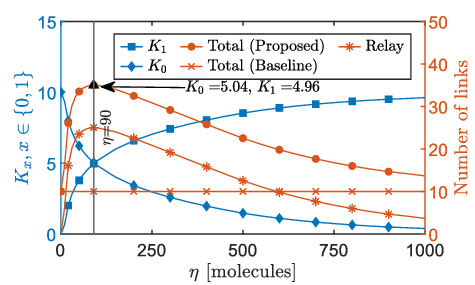}
\caption{Left $y$-axis: $K_1$ and $K_0$ versus $\eta$. Right $y$-axis: Number of total and relay links versus $\eta$.}
\label{fig_KF_Paths_vs_eta_Rec_10} \vspace{-.4cm}
\end{figure}

In Fig. \ref{fig_KF_Paths_vs_eta_Rec_10}, we analyze the impact of varying $\eta$ on the distribution of $K_0$ and $K_1$, as well as that of the total and relay links. As $\eta$ increases, DgNs with relatively better channel conditions to the infected tissue $S$ (i.e., higher CIR) are included in cluster $\mathcal{K}_1$. Consequently, $K_1$ increases and $K_0$ decreases, demonstrating that our proposed scheme effectively localizes DgNs without requiring explicit proximity information. Additionally, the number of total and relay links fluctuates with $\eta$, peaking at $\eta=90$ when the ratio of $K_0$ to $K_1$ is approximately 1:1. This is expected as all DgNs in subsequent clusters can serve each DgN in a previous cluster.

\begin{figure}  
\centering
  \includegraphics[width=0.9\columnwidth]{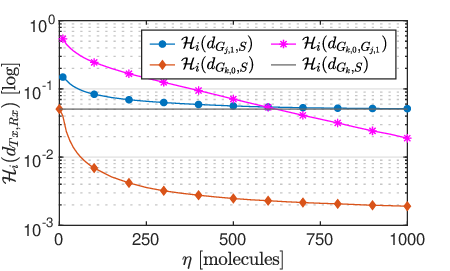}
  \caption{CIR of DgNs $G_{k,0}$/$G_{j,1}$ to the infected tissue and between $G_{k,0}$ and $G_{j,1}$ versus $\eta$. } 
\label{fig_CIR_eta_Rec_10} \vspace{-.4cm}
\end{figure}

In Fig. \ref{fig_CIR_eta_Rec_10}, we present the impact of varying $\eta$ on the CIRs across different links. As $\eta$ increases and more DgNs are included in cluster $\mathcal{K}_1$, the CIR between DgNs $G_{k,0}$ and $S$ (orange curve) and between DgNs $G_{k,0}$ and $G_{j,1}$ (magenta curve) declines. This is due to the MX channel, where more DgNs absorbing molecules before they reach $S$ reduces the CIR. Additionally, the CIR between DgNs $G_{j,1}$ and $S$ declines as more DgNs with weaker CIRs join the cluster. This illustrates that as the number of relays increases, the system's CIR diminishes, emphasizing the importance of optimizing relay configuration to balance the trade-offs. Lastly, when the number of DgNs in each cluster equals $K$, i.e., relaying is absent, the CIR matches the baseline system without relays (black curve).

\begin{figure}  
\centering
  \includegraphics[width=0.9\columnwidth]{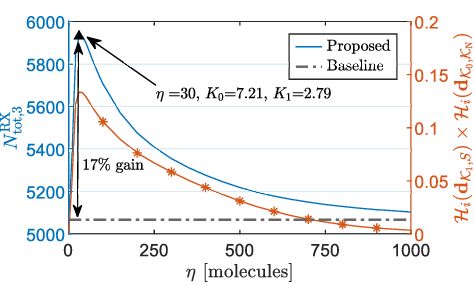}
  \caption{Left $y$-axis:$ N_{\text{tot},3}^\text{RX}$ versus $\eta$. Black triangle denote maximum $ N_{\text{tot},3}^\text{RX}$. Right $y$-axis: Total CIR including between DgNs in $\mathcal{K}_0$ and $\mathcal{K}_1$ and between DgNs in $\mathcal{K}_1$ and $S$.} 
\label{fig_DrugsDel_vs_eta_Rec_10}\vspace{-.3cm}
\end{figure}

In Fig. \ref{fig_DrugsDel_vs_eta_Rec_10}, the proposed scheme's efficiency is evident, delivering $17\%$ more drug molecules than the baseline without increasing drug quantity or dosage. The fluctuations in the number of drug molecules delivered resemble the variations in the number of links. However, the peak drug delivery occurs at $\eta =30$, rather than at $\eta =90$, as marked by the black triangle. This suggests that CIR, rather than the number of links, is critical in optimizing drug delivery. Therefore, optimal drug delivery depends not only on maximizing relay links but also on optimizing their CIR to ensure efficient delivery to the infected tissue. This finding reinforces the idea that reducing the distance between DgNs and the infected tissue is crucial for improving the performance of MC-TDD systems. 
  
\section{Conclusion}\label{sec_conclusion}
This study introduces a localization-enabled relaying mechanism that significantly improves the drug delivery efficiency of MC-TDD systems. A key finding is that optimizing the CIR is more effective than merely increasing the number of relays, emphasizing the importance of balancing relay configuration with delivery efficiency. This insight paves the way for further exploration into CIR-based relay selection as an optimization problem. Future research should focus on examining factors such as drug molecule lifespan, dynamic mobility of DgNs, and environmental influences, which are essential for advancing the precision and effectiveness of targeted drug delivery, especially in challenging contexts like localized infections and tumors.

\section{Acknowledgments}

This study was partly supported by the EU's Horizon Europe research and innovation programme under the Marie Skłodowska-Curie grant agreement No 101154851 and partly by the National Research Foundation of Korea (NRF) grant funded by the Korean Government (MSIT) (2021R1A2C1003507).
\bibliographystyle{ACM-Reference-Format}
\bibliography{REFS_LocRelay}


\begin{thebibliography}{22}


\ifx \showCODEN    \undefined \def \showCODEN     #1{\unskip}     \fi
\ifx \showDOI      \undefined \def \showDOI       #1{#1}\fi
\ifx \showISBNx    \undefined \def \showISBNx     #1{\unskip}     \fi
\ifx \showISBNxiii \undefined \def \showISBNxiii  #1{\unskip}     \fi
\ifx \showISSN     \undefined \def \showISSN      #1{\unskip}     \fi
\ifx \showLCCN     \undefined \def \showLCCN      #1{\unskip}     \fi
\ifx \shownote     \undefined \def \shownote      #1{#1}          \fi
\ifx \showarticletitle \undefined \def \showarticletitle #1{#1}   \fi
\ifx \showURL      \undefined \def \showURL       {\relax}        \fi
\providecommand\bibfield[2]{#2}
\providecommand\bibinfo[2]{#2}
\providecommand\natexlab[1]{#1}
\providecommand\showeprint[2][]{arXiv:#2}

\bibitem[Akyildiz et~al\mbox{.}(2008)]%
        {akyildiz2008nanonetworks}
\bibfield{author}{\bibinfo{person}{Ian~F. Akyildiz}, \bibinfo{person}{Fernando Brunetti}, {and} \bibinfo{person}{Cristina Bl\'{a}zquez}.} \bibinfo{year}{2008}\natexlab{}.
\newblock \showarticletitle{Nanonetworks: A New Communication Paradigm}.
\newblock \bibinfo{journal}{\emph{Elsevier Computer Networks}} \bibinfo{volume}{52}, \bibinfo{number}{12} (\bibinfo{date}{Aug.} \bibinfo{year}{2008}), \bibinfo{pages}{2260--2279}.
\newblock


\bibitem[Cheng et~al\mbox{.}(2022)]%
        {Cheng2022JointOptimizations}
\bibfield{author}{\bibinfo{person}{Zhen Cheng}, \bibinfo{person}{Jun Yan}, \bibinfo{person}{Jie Sun}, \bibinfo{person}{Yuchun Tu}, {and} \bibinfo{person}{Kaikai Chi}.} \bibinfo{year}{2022}\natexlab{}.
\newblock \showarticletitle{Joint Optimizations of Relays Locations and Decision Threshold for Multi-Hop Diffusive Mobile Molecular Communication With Drift}.
\newblock \bibinfo{journal}{\emph{IEEE Transactions on NanoBioscience}} \bibinfo{volume}{21}, \bibinfo{number}{3} (\bibinfo{year}{2022}), \bibinfo{pages}{454--465}.
\newblock
\urldef\tempurl%
\url{https://doi.org/10.1109/TNB.2022.3156633}
\showDOI{\tempurl}


\bibitem[{Chude-Okonkwo} et~al\mbox{.}(2017)]%
        {Chude2017SurveyMCTDDD}
\bibfield{author}{\bibinfo{person}{U.~A.~K. {Chude-Okonkwo}}, \bibinfo{person}{R. {Malekian}}, \bibinfo{person}{B.~T. {Maharaj}}, {and} \bibinfo{person}{A.~V. {Vasilakos}}.} \bibinfo{year}{2017}\natexlab{}.
\newblock \showarticletitle{{Molecular Communication and Nanonetwork for Targeted Drug Delivery: A Survey}}.
\newblock \bibinfo{journal}{\emph{IEEE Communications Surveys and Tutorials}} \bibinfo{volume}{19}, \bibinfo{number}{4} (\bibinfo{year}{2017}), \bibinfo{pages}{3046--3096}.
\newblock


\bibitem[Felicetti et~al\mbox{.}(2016)]%
        {Felicetti2016Applications}
\bibfield{author}{\bibinfo{person}{L. Felicetti}, \bibinfo{person}{M. Femminella}, \bibinfo{person}{G. Reali}, {and} \bibinfo{person}{P. Liò}.} \bibinfo{year}{2016}\natexlab{}.
\newblock \showarticletitle{{Applications of molecular communications to medicine: A survey}}.
\newblock \bibinfo{journal}{\emph{Nano Communication Networks}}  \bibinfo{volume}{7} (\bibinfo{year}{2016}), \bibinfo{pages}{27--45}.
\newblock
\showISSN{1878-7789}
\urldef\tempurl%
\url{https://doi.org/10.1016/j.nancom.2015.08.004}
\showDOI{\tempurl}


\bibitem[Femminella et~al\mbox{.}(2015)]%
        {Femminella2015MCDrugDel}
\bibfield{author}{\bibinfo{person}{Mauro Femminella}, \bibinfo{person}{Gianluca Reali}, {and} \bibinfo{person}{Athanasios~V. Vasilakos*}.} \bibinfo{year}{2015}\natexlab{}.
\newblock \showarticletitle{{A Molecular Communications Model for Drug Delivery}}.
\newblock \bibinfo{journal}{\emph{IEEE Transactions on NanoBioscience}} \bibinfo{volume}{14}, \bibinfo{number}{8} (\bibinfo{year}{2015}), \bibinfo{pages}{935--945}.
\newblock
\urldef\tempurl%
\url{https://doi.org/10.1109/TNB.2015.2489565}
\showDOI{\tempurl}


\bibitem[Guengerich(2011)]%
        {Guengerich2011Toxicity}
\bibfield{author}{\bibinfo{person}{F~Peter Guengerich}.} \bibinfo{year}{2011}\natexlab{}.
\newblock \showarticletitle{{Mechanisms of drug toxicity and relevance to pharmaceutical development}}.
\newblock \bibinfo{journal}{\emph{Drug metabolism and pharmacokinetics}} (\bibinfo{year}{2011}).
\newblock
\urldef\tempurl%
\url{https://www.ncbi.nlm.nih.gov/pmc/articles/PMC4707670/}
\showURL{%
\tempurl}


\bibitem[Guo et~al\mbox{.}(2021)]%
        {Guo2021}
\bibfield{author}{\bibinfo{person}{Weisi Guo}, \bibinfo{person}{Mahmoud Abbaszadeh}, \bibinfo{person}{Lin Lin}, \bibinfo{person}{Jerome Charmet}, \bibinfo{person}{Peter Thomas}, \bibinfo{person}{Zhuangkun Wei}, \bibinfo{person}{Bin Li}, {and} \bibinfo{person}{Chenglin Zhao}.} \bibinfo{year}{2021}\natexlab{}.
\newblock \showarticletitle{{Molecular Physical Layer for 6G in Wave-Denied Environments}}.
\newblock \bibinfo{journal}{\emph{IEEE Communications Magazine}} \bibinfo{volume}{59}, \bibinfo{number}{5} (\bibinfo{year}{2021}), \bibinfo{pages}{33--39}.
\newblock
\urldef\tempurl%
\url{https://doi.org/10.1109/MCOM.001.2000958}
\showDOI{\tempurl}


\bibitem[Jamali et~al\mbox{.}(2019)]%
        {Jamali2019ChannelTut}
\bibfield{author}{\bibinfo{person}{Vahid Jamali}, \bibinfo{person}{Arman Ahmadzadeh}, \bibinfo{person}{Wayan Wicke}, \bibinfo{person}{Adam Noel}, {and} \bibinfo{person}{Robert Schober}.} \bibinfo{year}{2019}\natexlab{}.
\newblock \showarticletitle{Channel Modeling for Diffusive Molecular Communication—A Tutorial Review}.
\newblock \bibinfo{journal}{\emph{Proc. IEEE}} \bibinfo{volume}{107}, \bibinfo{number}{7} (\bibinfo{year}{2019}), \bibinfo{pages}{1256--1301}.
\newblock
\urldef\tempurl%
\url{https://doi.org/10.1109/JPROC.2019.2919455}
\showDOI{\tempurl}


\bibitem[Kim and Chae(2013)]%
        {Kim2013NovelMod}
\bibfield{author}{\bibinfo{person}{Na-Rae Kim} {and} \bibinfo{person}{Chan-Byoung Chae}.} \bibinfo{year}{2013}\natexlab{}.
\newblock \showarticletitle{Novel Modulation Techniques using Isomers as Messenger Molecules for Nano Communication Networks via Diffusion}.
\newblock \bibinfo{journal}{\emph{IEEE Journal on Selected Areas in Communications}} \bibinfo{volume}{31}, \bibinfo{number}{12} (\bibinfo{year}{2013}), \bibinfo{pages}{847--856}.
\newblock
\urldef\tempurl%
\url{https://doi.org/10.1109/JSAC.2013.SUP2.12130017}
\showDOI{\tempurl}


\bibitem[Llatser et~al\mbox{.}(2012)]%
        {Llatser2012GrapheneAnt}
\bibfield{author}{\bibinfo{person}{Ignacio Llatser}, \bibinfo{person}{Christian Kremers}, \bibinfo{person}{Dmitry~N. Chigrin}, \bibinfo{person}{Josep~Miquel Jornet}, \bibinfo{person}{Max~C. Lemme}, \bibinfo{person}{Albert Cabellos-Aparicio}, {and} \bibinfo{person}{Eduard Alarcón}.} \bibinfo{year}{2012}\natexlab{}.
\newblock \showarticletitle{Characterization of graphene-based nano-antennas in the terahertz band}. In \bibinfo{booktitle}{\emph{2012 6th European Conference on Antennas and Propagation (EUCAP)}}. \bibinfo{pages}{194--198}.
\newblock
\urldef\tempurl%
\url{https://doi.org/10.1109/EuCAP.2012.6206598}
\showDOI{\tempurl}


\bibitem[Moritani et~al\mbox{.}(2005)]%
        {nakano2005molecular}
\bibfield{author}{\bibinfo{person}{Yuki Moritani}, \bibinfo{person}{Satoshi Hiyama}, \bibinfo{person}{Tatsuya Suda}, \bibinfo{person}{Ryota Egashira}, \bibinfo{person}{Akihiro Enomoto}, \bibinfo{person}{Michael Moore}, {and} \bibinfo{person}{Tadashi Nakano}.} \bibinfo{year}{2005}\natexlab{}.
\newblock \showarticletitle{{Molecular Communications between Nanomachines}}. In \bibinfo{booktitle}{\emph{24th IEEE Conference on Computer Communications (IEEE INFOCOM 2005)}}.
\newblock


\bibitem[Noel et~al\mbox{.}(2016)]%
        {Noel2016ChannelIR}
\bibfield{author}{\bibinfo{person}{Adam Noel}, \bibinfo{person}{Dimitrios Makrakis}, {and} \bibinfo{person}{Abdelhakim~Senhaji Hafid}.} \bibinfo{year}{2016}\natexlab{}.
\newblock \showarticletitle{Channel Impulse Responses in Diffusive Molecular Communication with Spherical Transmitters}.
\newblock \bibinfo{journal}{\emph{ArXiv}}  \bibinfo{volume}{abs/1604.04684} (\bibinfo{year}{2016}).
\newblock


\bibitem[Schulten(2015)]%
        {schulten2015lectures}
\bibfield{author}{\bibinfo{person}{K. Schulten}.} \bibinfo{year}{2015}\natexlab{}.
\newblock \bibinfo{booktitle}{\emph{Lectures in Theoretical Biophysics}}.
\newblock \bibinfo{publisher}{CreateSpace Independent Publishing Platform}.
\newblock
\showISBNx{9781506171715}
\urldef\tempurl%
\url{https://books.google.co.kr/books?id=sabwrQEACAAJ}
\showURL{%
\tempurl}


\bibitem[Shitiri and Cho(2021)]%
        {Shitiri2021Timing}
\bibfield{author}{\bibinfo{person}{Ethungshan Shitiri} {and} \bibinfo{person}{Ho-Shin Cho}.} \bibinfo{year}{2021}\natexlab{}.
\newblock \showarticletitle{{Timing Alignment in Molecular-Communication-Based Nanonetworks}}.
\newblock \bibinfo{journal}{\emph{IEEE Communications Magazine}} \bibinfo{volume}{59}, \bibinfo{number}{5} (\bibinfo{year}{2021}), \bibinfo{pages}{54--60}.
\newblock
\urldef\tempurl%
\url{https://doi.org/10.1109/MCOM.001.2000959}
\showDOI{\tempurl}


\bibitem[Shitiri et~al\mbox{.}(2021)]%
        {Shitiri2021Probability}
\bibfield{author}{\bibinfo{person}{Ethungshan Shitiri}, \bibinfo{person}{H.~Birkan Yilmaz}, {and} \bibinfo{person}{Ho-Shin Cho}.} \bibinfo{year}{2021}\natexlab{}.
\newblock \showarticletitle{Probability Distribution of a Signal’s Peak Time in a Molecular Diffusive Media}.
\newblock \bibinfo{journal}{\emph{IEEE Communications Letters}} \bibinfo{volume}{25}, \bibinfo{number}{12} (\bibinfo{year}{2021}), \bibinfo{pages}{3833--3837}.
\newblock
\urldef\tempurl%
\url{https://doi.org/10.1109/LCOMM.2021.3115724}
\showDOI{\tempurl}


\bibitem[Tewabe et~al\mbox{.}(2021)]%
        {tewabe2021TDD}
\bibfield{author}{\bibinfo{person}{Ashagrachew Tewabe}, \bibinfo{person}{Atlaw Abate}, \bibinfo{person}{Manaye Tamrie}, \bibinfo{person}{Abyou Seyfu}, {and} \bibinfo{person}{Ebrahim Abdela~Siraj}.} \bibinfo{year}{2021}\natexlab{}.
\newblock \bibinfo{title}{{Targeted drug delivery - from Magic Bullet to Nanomedicine: Principles, challenges, and future perspectives}}.
\newblock
\newblock
\urldef\tempurl%
\url{https://www.ncbi.nlm.nih.gov/pmc/articles/PMC8275483/}
\showURL{%
\tempurl}


\bibitem[Wang et~al\mbox{.}(2015)]%
        {Wang2015Relay}
\bibfield{author}{\bibinfo{person}{Xiayang Wang}, \bibinfo{person}{Matthew~D. Higgins}, {and} \bibinfo{person}{Mark~S. Leeson}.} \bibinfo{year}{2015}\natexlab{}.
\newblock \showarticletitle{Relay Analysis in Molecular Communications With Time-Dependent Concentration}.
\newblock \bibinfo{journal}{\emph{IEEE Communications Letters}} \bibinfo{volume}{19}, \bibinfo{number}{11} (\bibinfo{year}{2015}), \bibinfo{pages}{1977--1980}.
\newblock
\urldef\tempurl%
\url{https://doi.org/10.1109/LCOMM.2015.2478780}
\showDOI{\tempurl}


\bibitem[Xia(2020)]%
        {xia2020DrugEfficacy}
\bibfield{author}{\bibinfo{person}{Xuhua Xia}.} \bibinfo{year}{2020}\natexlab{}.
\newblock \bibinfo{title}{{Drug efficacy and toxicity prediction: An innovative application of Transcriptomic Data}}.
\newblock
\newblock
\urldef\tempurl%
\url{https://www.ncbi.nlm.nih.gov/pmc/articles/PMC7661398/}
\showURL{%
\tempurl}


\bibitem[Yaylali et~al\mbox{.}(2023)]%
        {yaylali2021channelMx}
\bibfield{author}{\bibinfo{person}{Gokberk Yaylali}, \bibinfo{person}{Bayram~Cevdet Akdeniz}, \bibinfo{person}{Tuna Tugcu}, {and} \bibinfo{person}{Ali~Emre Pusane}.} \bibinfo{year}{2023}\natexlab{}.
\newblock \showarticletitle{Channel Modeling for Multi-Receiver Molecular Communication Systems}.
\newblock \bibinfo{journal}{\emph{IEEE Transactions on Communications}} \bibinfo{volume}{71}, \bibinfo{number}{8} (\bibinfo{year}{2023}), \bibinfo{pages}{4499--4512}.
\newblock
\urldef\tempurl%
\url{https://doi.org/10.1109/TCOMM.2023.3281415}
\showDOI{\tempurl}


\bibitem[Yetimoglu et~al\mbox{.}(2022)]%
        {Oguzhan2022Multiple}
\bibfield{author}{\bibinfo{person}{Oguzhan Yetimoglu}, \bibinfo{person}{M.~Kerem Avci}, \bibinfo{person}{Bayram~Cevdet Akdeniz}, \bibinfo{person}{H.~Birkan Yilmaz}, \bibinfo{person}{Ali~E. Pusane}, {and} \bibinfo{person}{Tuna Tugcu}.} \bibinfo{year}{2022}\natexlab{}.
\newblock \showarticletitle{Multiple transmitter localization via single receiver in 3-D molecular communication via diffusion}.
\newblock \bibinfo{journal}{\emph{Digital Signal Processing}}  \bibinfo{volume}{124} (\bibinfo{year}{2022}), \bibinfo{pages}{103185}.
\newblock
\showISSN{1051-2004}
\urldef\tempurl%
\url{https://doi.org/10.1016/j.dsp.2021.103185}
\showDOI{\tempurl}
\newblock
\shownote{Signal Processing Aspects of Molecular Communications}.


\bibitem[{Yilmaz} et~al\mbox{.}(2014)]%
        {yilmaz2014Three}
\bibfield{author}{\bibinfo{person}{H.~B. {Yilmaz}}, \bibinfo{person}{A.~C. {Heren}}, \bibinfo{person}{T. {Tugcu}}, {and} \bibinfo{person}{C. {Chae}}.} \bibinfo{year}{2014}\natexlab{}.
\newblock \showarticletitle{{Three-Dimensional Channel Characteristics for Molecular Communications With an Absorbing Receiver}}.
\newblock \bibinfo{journal}{\emph{IEEE Communications Letters}} \bibinfo{volume}{18}, \bibinfo{number}{6} (\bibinfo{date}{June} \bibinfo{year}{2014}), \bibinfo{pages}{929--932}.
\newblock
\showISSN{1089-7798}
\urldef\tempurl%
\url{https://doi.org/10.1109/LCOMM.2014.2320917}
\showDOI{\tempurl}


\bibitem[Şükrü Kuran et~al\mbox{.}(2010)]%
        {Kuran2010Energy}
\bibfield{author}{\bibinfo{person}{Mehmet Şükrü Kuran}, \bibinfo{person}{H.~Birkan Yilmaz}, \bibinfo{person}{Tuna Tugcu}, {and} \bibinfo{person}{Bilge Özerman}.} \bibinfo{year}{2010}\natexlab{}.
\newblock \showarticletitle{{Energy Model for Communication via Diffusion in Nanonetworks}}.
\newblock \bibinfo{journal}{\emph{Nano Communication Networks}} \bibinfo{volume}{1}, \bibinfo{number}{2} (\bibinfo{year}{2010}), \bibinfo{pages}{86--95}.
\newblock
\showISSN{1878-7789}
\urldef\tempurl%
\url{https://doi.org/10.1016/j.nancom.2010.07.002}
\showDOI{\tempurl}


\end{thebibliography}
\appendix

\end{document}